\documentclass[journal,a4paper,onecolumn]{IEEEtran}

\usepackage{tikz}
\usepackage{epsfig}
\usepackage{epstopdf}
\usepackage{siunitx}
\usepackage{pbox}
\usepackage{makecell}
\usepackage{multirow}
\usepackage{amssymb}
\usepackage{amsmath}
\usepackage[normalem]{ulem}

\usepackage{blindtext}
\usepackage{etoolbox}
 
\makeatletter

\docsvlist{\@oddhead,\@evenhead,\ps@headings,\ps@IEEEtitlepagestyle,\ps@IEEEpeerreviewcoverpagestyle}
\makeatother

\begin{document}


\title{Experimental observation and modeling of the impact of traps on static and analog/HF performance of graphene transistors}

\author{Anibal Pacheco-Sanchez, Nikolaos Mavredakis, Pedro C. Feijoo, Wei Wei, \\Emiliano Pallecchi, Henri Happy, David Jiménez 

\thanks{This work has received funding from the European Union’s Horizon 2020 research and innovation programme under grant agreements No GrapheneCore2 785219 and No GrapheneCore3 881603, from Ministerio de Ciencia, Innovación y Universidades under grant agreement RTI2018-097876-B-C21(MCIU/AEI/FEDER, UE). This  article  has been partially  funded  by  the  European  Regional  Development  Funds  (ERDF)  allocated  to  the  Programa Operatiu FEDER de Catalunya 2014-2020, with the support of the Secretaria d’Universitats i Recerca of the Departament d’Empresa i Coneixement of the Generalitat de Catalunya for emerging technology clusters to  carry  out  valorization  and  transfer  of  research  results.  Reference  of  the  GraphCAT  project:  001-P-001702.\newline \indent A. Pacheco-Sanchez, N. Mavredakis, P. C. Feijoo and D. Jiménez are with the Departament d'Enginyeria Electr\`{o}nica, Escola d'Enginyeria, Universitat Aut\`{o}noma de Barcelona, Bellaterra 08193, Spain (e-mails: AnibalUriel.Pacheco@uab.cat, Nikolaos.Mavredakis@uab.cat).\newline\indent W. Wei, E. Pallecchi and H. Happy are with CNRS, UMR 8520—IEMN, University of Lille, 59000 Lille, France.}
}
\maketitle

\makeatletter
\def\ps@IEEEtitlepagestyle{
  \def\@oddfoot{\mycopyrightnotice}
  \def\@evenfoot{}
}
\def\mycopyrightnotice{
  {\footnotesize
  \begin{minipage}{\textwidth}
  \centering
© 2020 IEEE.  Personal use of this material is permitted.  Permission from IEEE must be obtained for all other uses, in any current or future media, including reprinting/republishing this material for advertising or promotional purposes, creating new collective works, for resale or redistribution to servers or lists, or reuse of any copyrighted component of this work in other works.
  \end{minipage}
  }
}

\begin{abstract}
\boldmath
The trap-induced hysteresis on the performance of a graphene field-effect transistor is experimentally diminished here by applying consecutive gate-to-source voltage pulses of opposing polarity. This measurement scheme is a practical and suitable approach to obtain reproducible device characteristics. Trap-affected and trap-reduced experimental data enable a discussion regarding the impact of traps on static and dynamic device performance. An analytical drain current model calibrated with the experimental data enables the study of the traps effects on the channel potential within the device. High-frequency figures of merit and the intrinsic gain of the device obtained from both experimental and synthetic data with and without hysteresis show the importance of considering the generally overlooked impact of traps for analog and high-frequency applications. 
 
\end{abstract}
%
%

\IEEEpeerreviewmaketitle
\section{Introduction}
\label{ch:intro}
%

Reproducible graphene field-effect transistor (GFETs) characteristics are required in order to boost the use of circuits based on this emerging technology, specially on the low-power high-frequency (HF) applications scenario where its extraordinary intrinsic characteristics, e.g., high carrier mobility, can be exploited \cite{Sch13}. In GFETs, traps within the channel, substrate and high-$\kappa$ oxide materials, as well as at the interfaces between them, are one of the major issues affecting the device performance, i.e., in order to obtain hysteresis-free reproducible characteristics, the impact of traps should be diminished \cite{CarSer14}, \cite{MukFre14}. Technological efforts towards reducing the presence of traps in graphene-based technologies have shown a discrete success on individual devices \cite{MukFre14}-\cite{NatZha19}. For wafer-scale integration however, defect-induced traps within the channel associated to the graphene transfer process, as well as oxide traps, still affect the device behavior \cite{SmiWag17}, \cite{NeuPin19}. The characterization of traps in GFETs is therefore required at this stage of the technology in order to enhance device and circuit reproducibility by finding the adequate bias conditions to diminish their impact on the overall device performance.

Trap mechanisms within GFET architectures have been experimentally characterized by observing the electrical device characteristics obtained with time-dependent-voltage pulses \cite{CarSer14}-\cite{NatZha19}, \cite{LeeKan11}-\cite{MisMee17}. Different time constants related to trapping and detrapping processes have been identified for different graphene technologies ranging over a large span of time, depending on the location of the traps (bulk, oxide, channel, interfaces) \cite{CarSer14}, \cite{LeeKan11}-\cite{RamSom17}. In most of the GFETs trap studies, the drain-to-source voltage $V_{\rm{DS}}$ signal has been varied over time while keeping the gate-to-source $V_{\rm{GS}}$ constant \cite{MukFre14}, \cite{NatZha19}, \cite{LeeKan11}, \cite{RamSom17}, \cite{MisMee17} i.e., $V_{\rm{DS}}$-induced hysteresis has been the main focus of such investigations rather than the trap impact on the device performance due to vertical fields applied to the channel. The latter effect has been studied for a global-back-gate device with SiO$_2$ dielectric \cite{MaoWan16} and for top-gate devices with high-$\kappa$ gate oxides \cite{CarSer14} in order to obtain the trapping time constants of such specific technologies. 

The impact of traps on analog/HF performance of GFETs has been rarely discussed. Some trapping processes can be too slow in comparison to the targeted operation frequency ($\sim$\SI{}{\giga\hertz}), i.e., their impact is generally neglected based on this assumption \cite{MadHol12}. However, the DC bias point, required to activate the transistor for its dynamic operation, is still affected by traps, i.e., the drift of this point due to traps can induce reproducibility issues in GFET-based HF applications, as demonstrated for other emerging transistor technologies \cite{HafPac16}, \cite{HelLin20}. HF figures of merit (FoM) have been reported for GFETs measured under pulsed bias conditions \cite{MukFre14}, \cite{SanXu18}, however, no trap-affected data have been shown in order to understand the difference in the dynamic performance with and without hysteresis. As an alternative to overcome the challenging measurement setup required for pulsed HF characterization \cite{MukFre14}, \cite{HafPac16}, the HF FoM of GFETs can be measured by considering a non-quiescent holding time large enough for the traps to be inactive \cite{MavGar18}-\cite{AsaBon20}. The latter is a useful approach to understand the internal device mechanisms after removing the impact of traps. However, such holding time for trap-reduced characterization varies between technologies and depends on the measurement history \cite{CarSer14}, \cite{LeeKan11}-\cite{RamSom17}. Furthermore, in a practical HF scenario, GFET-based circuits should work under rapid pulsed biasing schemes rather than on a holding time-based scheme in order to compete with incumbent technologies, e.g., in applications involving high-data rate communications with different pulse-based modulation schemes \cite{LeeLee12}, \cite{HabHe17}. Hence, an alternative practical biasing method to achieve trap-reduced GFET performance for its use in HF applications is required. 

The aims of this work are to show experimentally the impact of traps on the GFET static and dynamic performance and to provide a straightforward model of trap-related effects. For the first goal, the trap-affected and trap-reduced performance of a buried-gate GFET technology are characterized with a standard $V_{\rm{GS}}$-staircase sweep and with an opposing-pulse sweep of $V_{\rm{GS}}$, respectively. The opposing-pulse sweep allows to determine the impact of traps on transistor performance indicators and it is a practical measurement approach to obtain reproducible characteristics. In second place, this paper successfully describes the impact of traps on the net charge within the device via an analytical compact model. Typical models only consider traps activated by vertical electrical fields caused by the gate voltage stress, but our model also include traps activated by hot-carriers, which result in a $V_{\rm{DS}}$-dependence of trapped charge. Our compact model thus allows calculation of traps densities and enables the study of GFET-based circuits considering devices affected by traps at different bias points. This work projects the dynamic performance via HF/analog FoM, including cutoff frequency, maximum oscillation frequency and intrinsic gain. Additionally, it provides an insight on the generally neglected fact that traps affect not only the static device performance but also the dynamic response at different bias.

\section{Device description and measurement techniques} \label{ch:DUT}

A two-finger aluminium back-gated GFET fabricated on a SiO$_2$ (\SI{300}{\nano\meter})/Si (high resistivity silicon substrate) has been characterized in this study. The device gate width $w_{\rm{g}}$ and gate length are $\SI{12}{}\times\SI{2}{\micro\meter}$ and \SI{300}{\nano\meter}, respectively (the transistor is a dual gate of \SI{12}{} microns each). The single graphene layer has been grown via chemical vapor deposition in a host substrate (a copper foil) and transferred on the top of previously patterned bottom Al gate fingers via a wet chemical transfer technique. The natural oxidation of the Al back-gate has been obtained prior the transfer of monolayer of graphene. A $\sim$\SI{4}{\nano\meter} thick Al$_2$O$_3$ dielectric layer (ellipsometry measurement) is then obtained which separates graphene and gate. The Ni/Au source and drain contacts are separated by a distance of \SI{1}{\micro\meter}. The gate is located at the same distance from source and drain contacts. More details on the graphene transfer process and on the device fabrication process as well as the device layout have been discussed elsewhere \cite{WeiZho15}.

\begin{figure}[!htb]
\centering
\includegraphics[height=0.2475\textwidth]{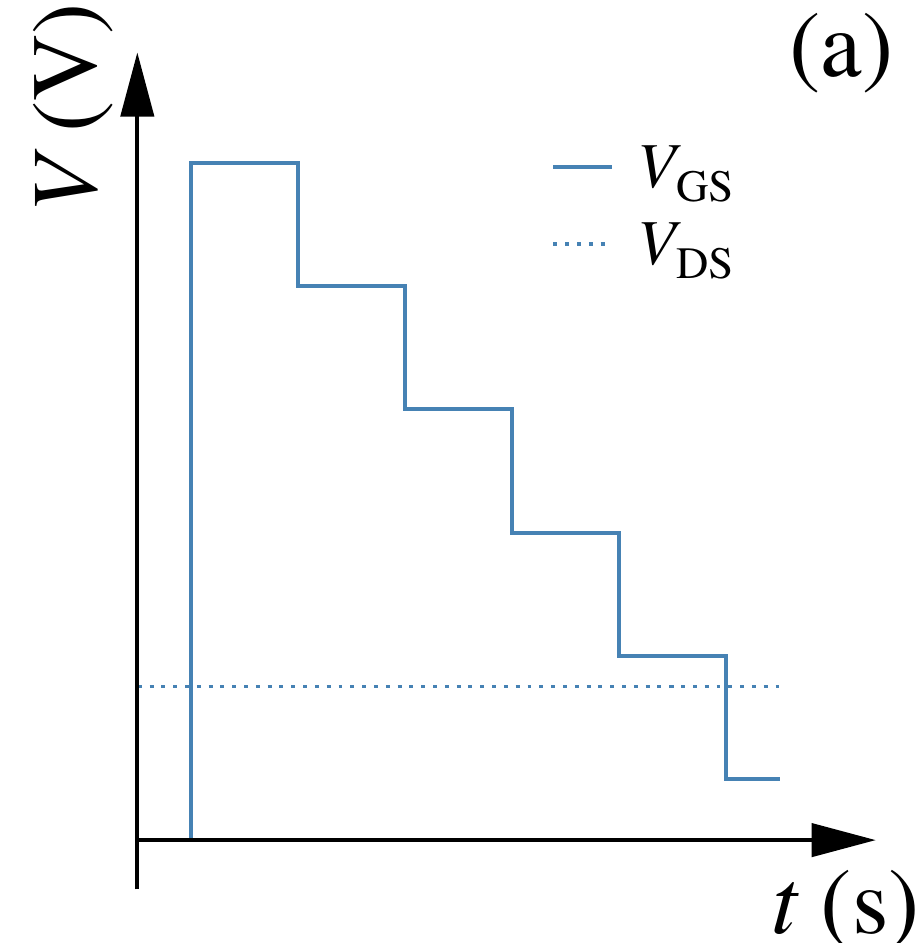}
\includegraphics[height=0.2475\textwidth]{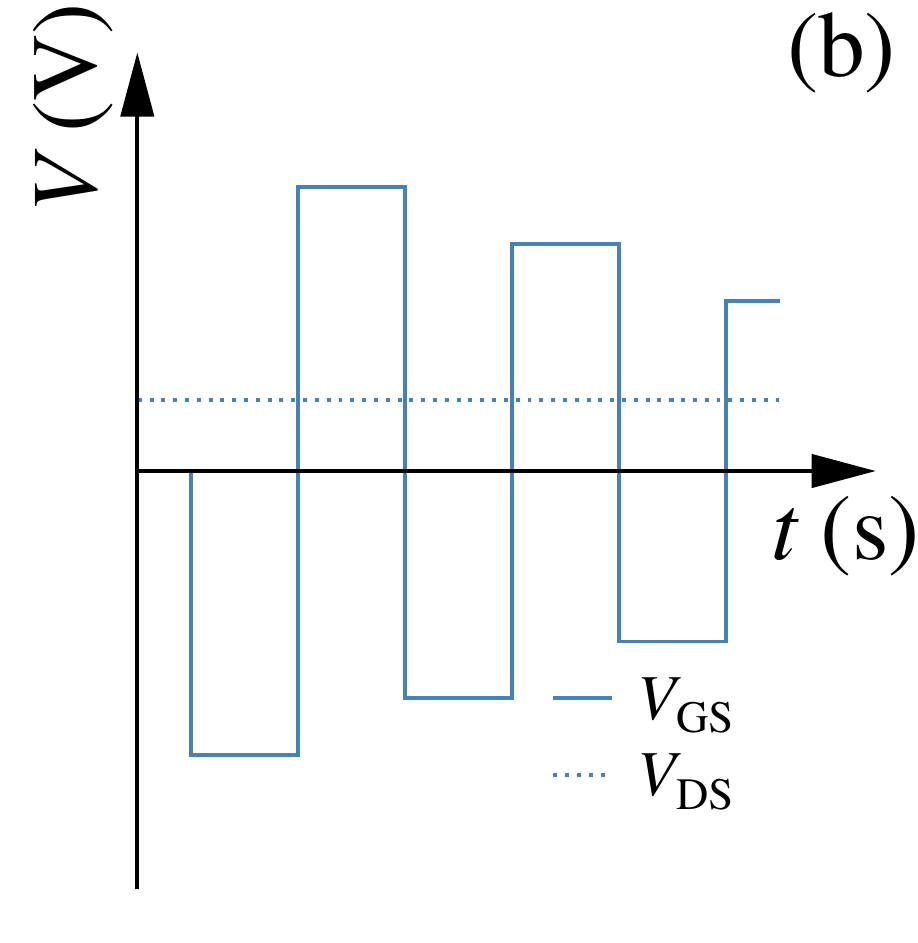} \\
\caption{Sketch of the applied voltage signals over time for (a) staircase sweep and (b) opposing pulses sweep measurements.}
\label{fig:sweeps}
\end{figure}

Two different measurement techniques have been used for the experimental device characterization: staircase voltage sweep and opposing voltage sweep. Fig. \ref{fig:sweeps}(a) shows a sketch of the $V_{\rm{GS}}$ and $V_{\rm{DS}}$ for the staircase sweep characterization. This standard characterization scheme usually acquires the current data at the end of the pulse where a steady-state current is expected. This steady-state in graphene devices however, depends on the technology and measurement history as elucidated by the wide span of trap-time constants reported in different studies (from \SI{}{\nano\second} to \SI{}{s}) \cite{CarSer14}, \cite{NatZha19}, \cite{LeeKan11}-\cite{MisMee17}, \cite{AsaBon20}. 

The opposing sweep technique sketched in Fig. \ref{fig:sweeps}(b) diminishes the trap-effects since the measurement history is compensated by the consecutive opposite bias pulses, i.e., in time-symmetric trapping and detrapping processes, the effects related to remaining carriers trapped at the end of a positive or negative $V_{\rm{GS}}$ pulse, if any, are counterbalanced by trapped carriers at the consecutive pulse of similar magnitude but opposite bias. This reversible trapping effects technique has been previously reported for devices with high-$\kappa$ oxides \cite{HafPac16}, \cite{ZafKum05}-\cite{PacBej19} but not exploited before for GFETs. Furthermore, this bias scheme is less challenging to implement in contrast to a pulsed characterization technique and more practical in circuit applications than a holding-time bias approach, e.g., in modulators required for high-data rate communications \cite{LeeLee12}, \cite{HabHe17}.

Measurements with both techniques described above have been performed with a K4200 semiconductor characterization system by applying consecutive forward and backward $V_{\rm{GS}}$-sweeps at room temperature. $V_{\rm{DS}}$ has been kept constant. The duration of the applied $V_{\rm{GS}}$ signal is of $\sim$\SI{0.9}{\second} which is large enough to consider trapping processes to be active according to time constants in the same range obtained for GFETs \cite{MaoWan16}, \cite{RamSom17}. This is confirmed by the hysteresis observed in the experimental section. Self-heating effects are expected to be much faster \cite{RamSom17}, \cite{MisMee17} than the trapping processes characterized with this pulse width. Notice that in contrast to other studies \cite{MukFre14}, \cite{NatZha19}, \cite{LeeKan11}, \cite{RamSom17}, \cite{MisMee17}, the gate oxide traps are directly affected by transitions of the applied vertical fields here.

\section{Experimental investigation} \label{ch:exp}

\begin{figure}[!htb]
\centering
\includegraphics[height=0.2475\textwidth]{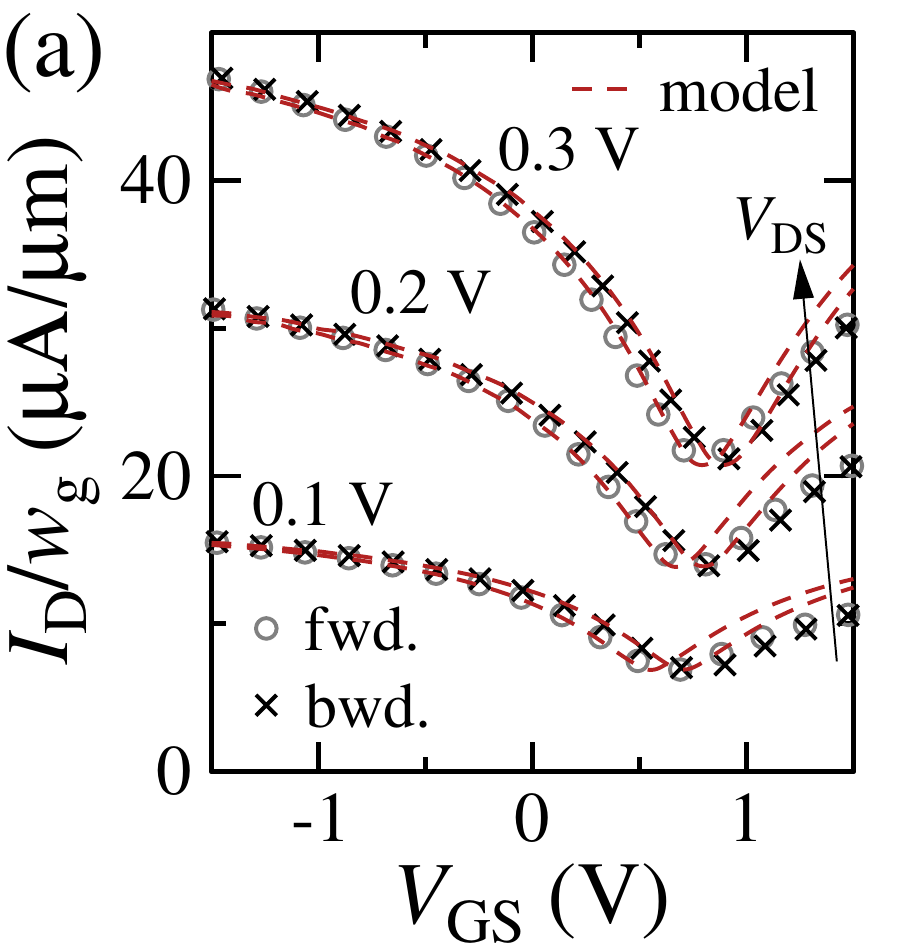}
\includegraphics[height=0.2475\textwidth]{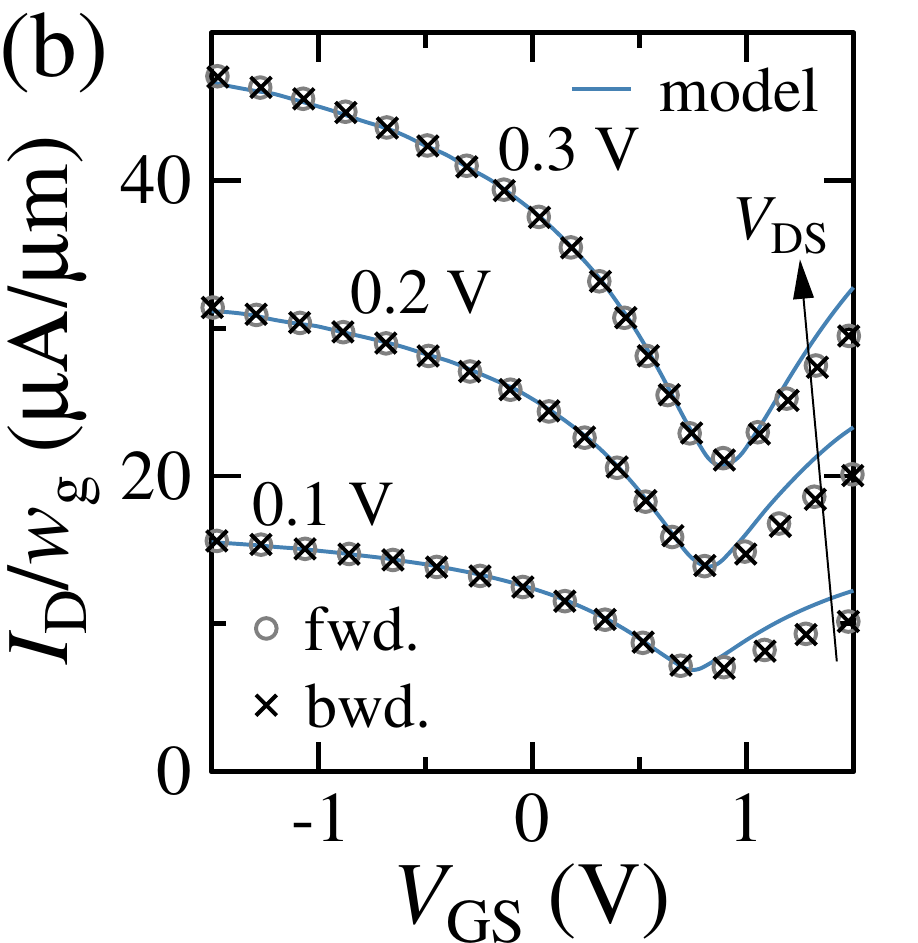} \\
\caption{Transfer characteristics of the \SI{300}{\nano\meter}-long GFET with (a) stair-case sweep and (b) opposing pulse sweep characterization. Markers are experimental data and lines represent modeling results.}
\label{fig:IdVg_all}
\end{figure}

The channel potential of the device can be modified by interface or oxide traps capturing or releasing carriers since these processes reduce the gate control, i.e., traps shield the channel potential from the applied voltage \cite{BonVor17}. These trap effects have a different impact on the device performance depending on the measurement technique. The transfer characteristics of the device obtained via consecutive forward-backward sweeps with the staircase and opposing measurement techniques are shown in Fig. \ref{fig:IdVg_all}. The typical ambipolar behavior of GFETs, separated by a charged neutrality voltage point identified as the Dirac voltage $V_{\rm{Dirac}}=V_{\rm{GS}}\vert_{\rm{min}\left(\mathit{I}_{\rm{D}}\right)}$, can be observed.

The hysteresis in the experimental characteristics measured with forward and backward sweeps using the staircase technique (Fig. \ref{fig:IdVg_all}(a)) is induced by trapping processes started during the forward sweep and continued during the change of sweep direction. This leads to lower $I_{\rm{D}}$ and higher $V_{\rm{Dirac}}$ in the backward sweep compared to the forward sweep at $V_{\rm{GS}}>V_{\rm{Dirac}}$. The release of trapped carriers at $V_{\rm{GS}}<V_{\rm{Dirac}}$ during the backward sweep affects positively the channel potential \cite{LeeKan11}, i.e., higher $I_{\rm{D}}$ is observed in contrast to the acquired during the former sweep. The reduced hysteresis observed at more negative $V_{\rm{GS}}$ can be associated to the device reaching a steady-like-state. 

\begin{figure}[!htb]
\centering
\includegraphics[height=0.2475\textwidth]{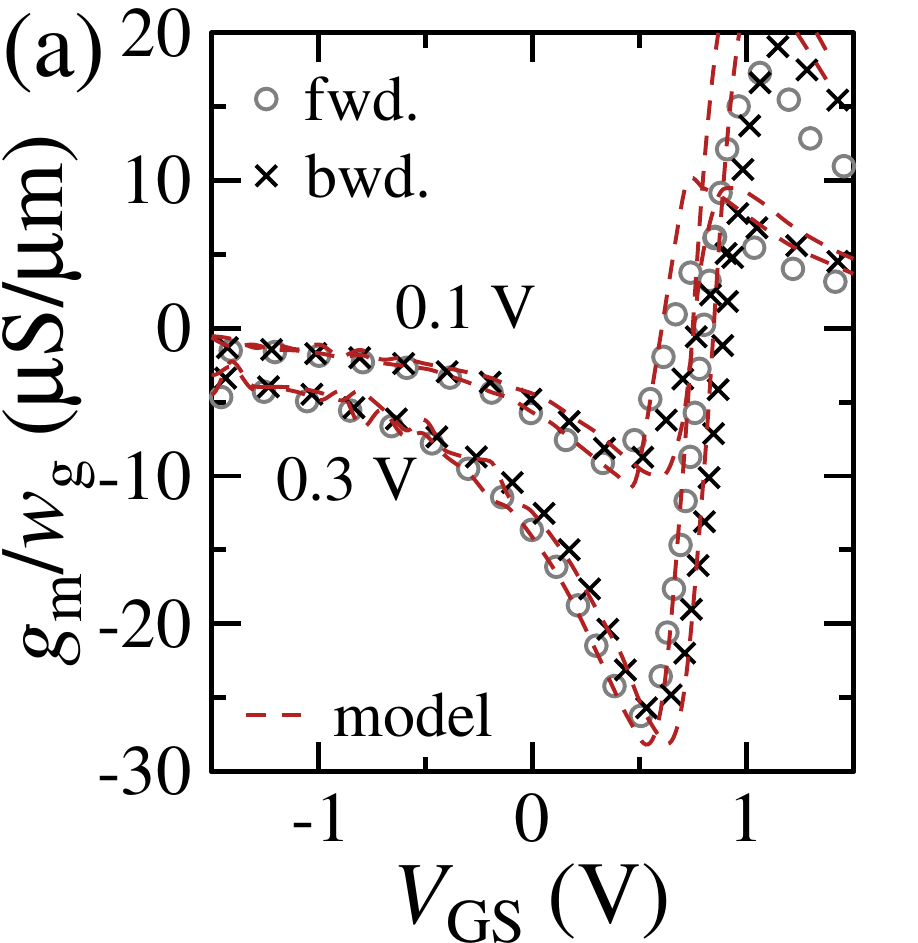}
\includegraphics[height=0.2475\textwidth]{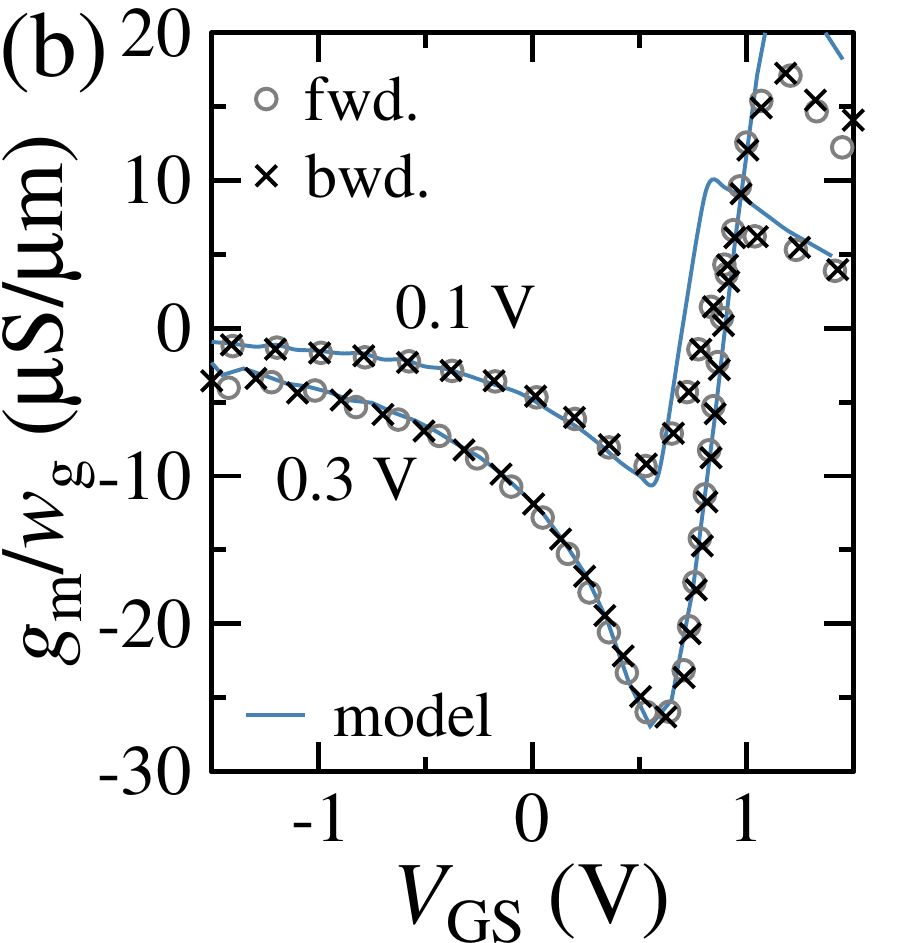} \\
\caption{Transconductance versus $V_{\rm{GS}}$ at different $V_{\rm{DS}}$ of a \SI{300}{\nano\meter}-long GFET with (a) stair-case sweep and (b) opposing pulse sweep characterization. Markers are experimental data and lines represent modeling results.}
\label{fig:gmVg_all}
\end{figure}

\begin{figure}[!htb]
\centering
\includegraphics[height=0.2475\textwidth]{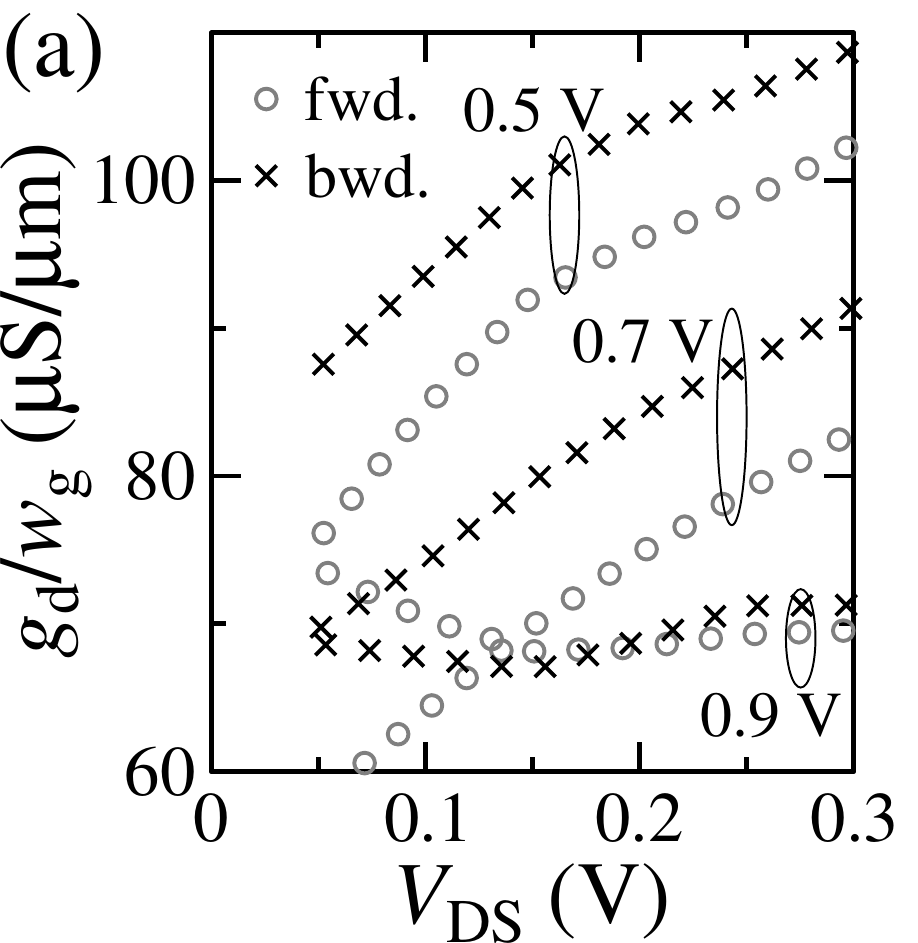}
\includegraphics[height=0.2475\textwidth]{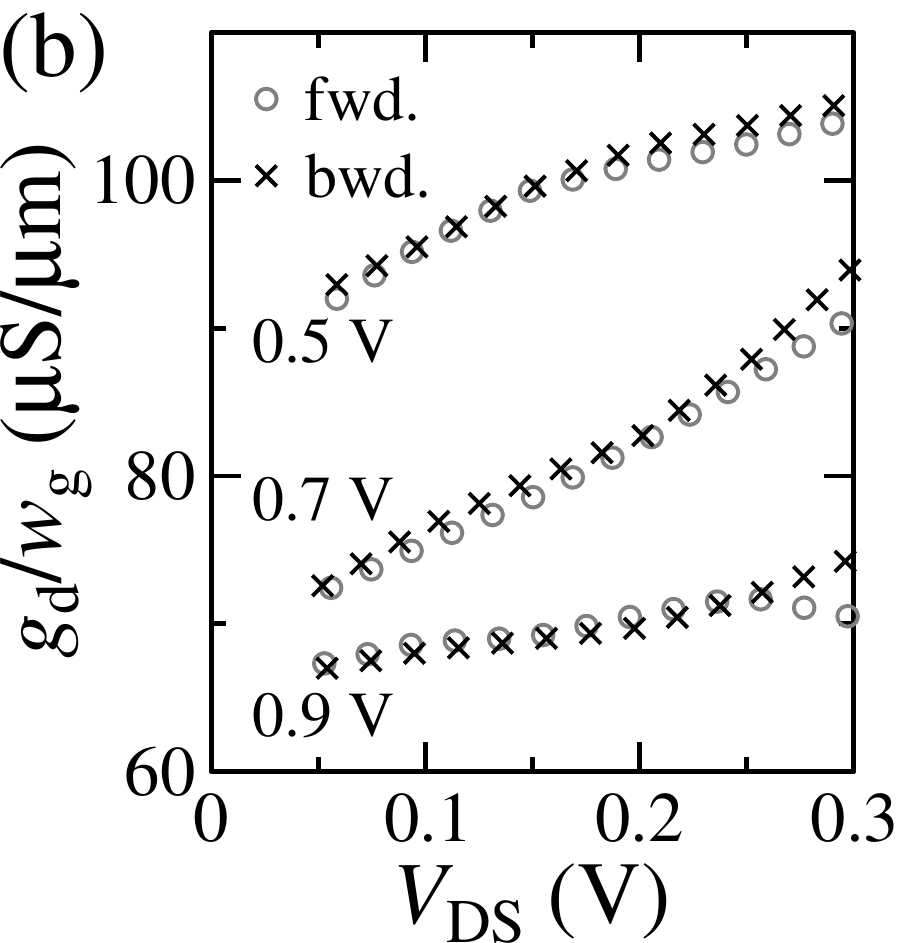} \\
\caption{Output conductance versus $V_{\rm{DS}}$ at different $V_{\rm{GS}}$ of a \SI{300}{\nano\meter}-long GFET with (a) stair-case sweep and (b) opposing pulse sweep characterization.}
\label{fig:gdVg_all}
\end{figure}

The almost negligible hysteresis observed in the transfer characteristic measured with the opposing pulse technique, as observed in Fig. \ref{fig:IdVg_all}(b), can be explained by trapping and detrapping processes with quasi-symmetrical time constants and by the compensating effect due to opposing polarity pulses as mentioned in Section \ref{ch:DUT}. The transconductance ($g_{\rm{m}}=\partial I_{\rm{D}} / \partial V_{\rm{GS}}$) and output conductance ($g_{\rm{d}}=\partial I_{\rm{D}} / \partial V_{\rm{DS}}$) plots in Figs. \ref{fig:gmVg_all} and \ref{fig:gdVg_all}, make  the hysteresis in the staircase sweeps more evident. The slight observed hysteresis in the opposing sweep data at high $V_{\rm{DS}}$ can be related to hot carriers-induced \cite{CarSer14} asymmetric trapping processes which can not be counterbalanced with this technique.

Transistor performance indicators oftenly used to project the device HF performance are the maximum point of the transconductance $g_{\rm{m,max}}$ and a low value of $g_{\rm{d}}$. Due to traps, however, the bias-point and the magnitude of each of these parameters differ in each measurement run. E.g., by analyzing the sweeps with the staircase technique, $\vert g_{\rm{m,max}} \vert$ is found at different $V_{\rm{GS}}$ and its magnitude differ in comparison to the same metric obtained with both sweeps using the opposing pulse technique. Furthermore,  a $\vert g_{\rm{m,max}} \vert$ equal to \SI{634}{\micro\siemens} at the highest $V_{\rm{DS}}$ obtained with trap-affected data is $\sim$\SI{21}{\micro\siemens} lower than the same metric observed at the same $V_{\rm{DS}}$ with trap-reduced data.  A large difference between trap-affected and trap-reduced data is observed for the output conductance (Fig. \ref{fig:gdVg_all}). In addition to the distinct values obtained in each case, the bias dependence of trap-affected data is notoriously different between different sweeps. Variations in the operating bias point and magnitude of $g_{\rm{m}}$ and $g_{\rm{d}}$ can mislead the expected analog/HF performance projections based on this parameter such as the maximum oscillation frequency.

\begin{figure}[!htb]
\centering
\includegraphics[height=0.2475\textwidth]{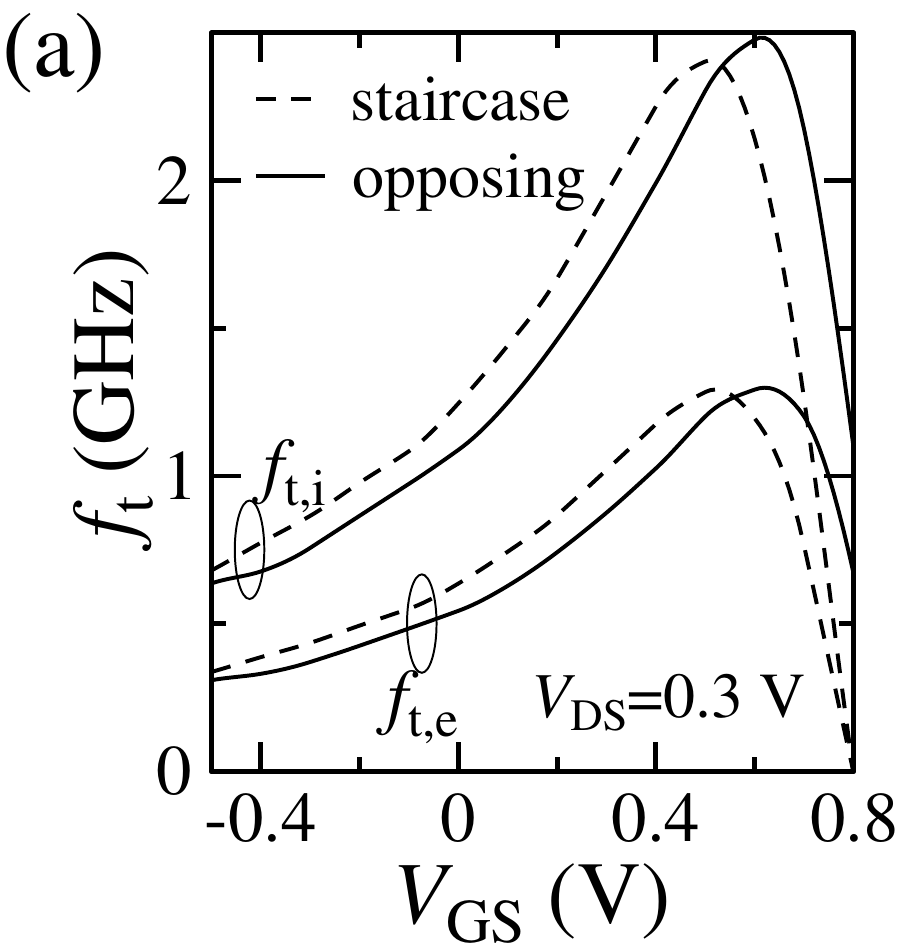}
\includegraphics[height=0.2475\textwidth]{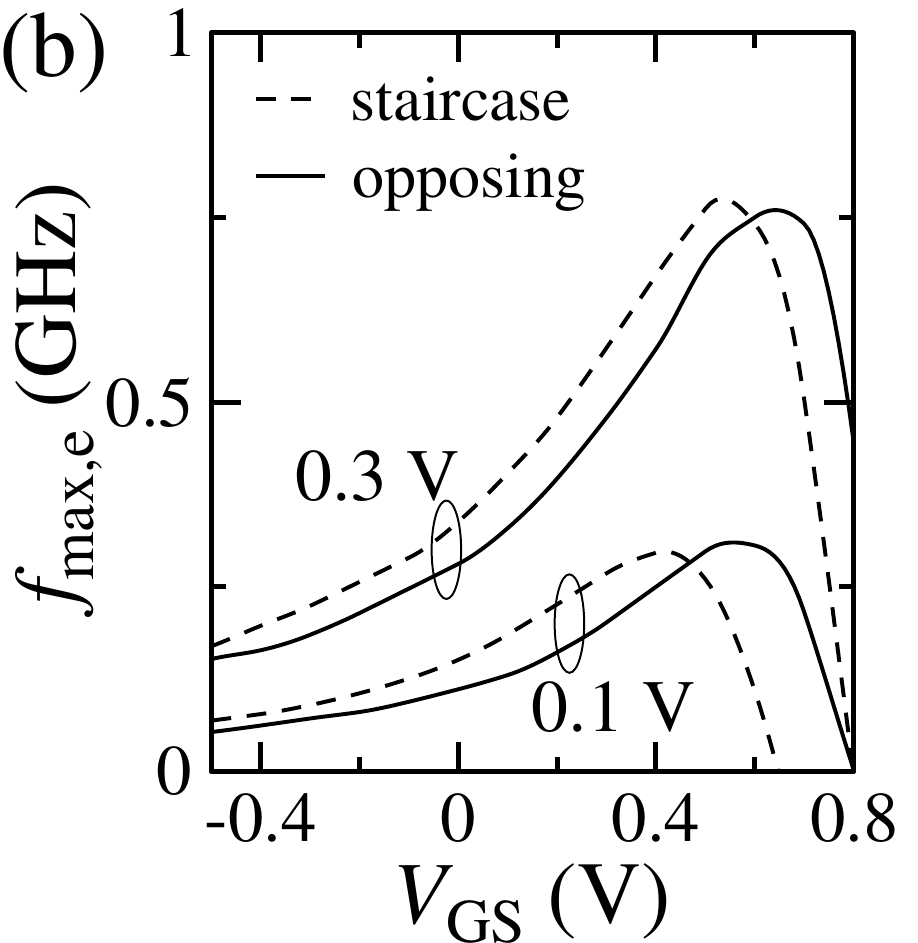} \\
\caption{Trap-affected (staircase sweep) and trap-reduced (opposing sweep) calculated (a) intrinsic and extrinsic cutoff frequency and (b) extrinsic maximum oscillation frequency versus $V_{\rm{GS}}$ of the device under study.}
\label{fig:ft}
\end{figure}

In order to explicitly show the impact of traps on analog/HF FoMs, the intrinsic/extrinsic cutoff frequency $f_{\rm{t,i/e}}$ and the extrinsic maximum oscillation frequency $f_{\rm{max,e}}$ have been estimated from the trap-affected and trap-reduced data of the device by an approximation based on a small-signal model of GFETs \cite{Sch13}, \cite{PacFei20}, \cite{GuoDon13}. An average gate capacitance of \SI{42}{\femto\farad} obtained elsewhere \cite{WeiZho15} for the device used here has been considered. An effective physical gate resistance \cite{VanGee94} of \SI{4.9}{\ohm} has been calculated. A contact resistance value (see Table \ref{tab:rc_exp}), required for the intrinsic transconductance and intrinsic output conductance, has been extracted from the experimental data with a method presented elsewhere \cite{PacFei20}. The obtained HF FoM have been shown in Fig. \ref{fig:ft} for the forward sweep of each technique. 

The magnitude and bias-dependence of the HF FoMs differ due to traps affecting differently the channel potential. The trap-induced bias drift and the value of $f_{\rm{t,e}}$ and $f_{\rm{max,e}}$ can affect critically GFET-based HF circuits performance, e.g., by misleading the design for a specific bias point of matching or stability networks connecting the device with other stages of a monolithic circuit \cite{YuHe16}. 

Other transistor performance indicators obtained with the different characterization techniques are shown in Table \ref{tab:rc_exp} at the highest $V_{\rm{DS}}$ used here. For comparison purposes, $V_{\rm{Dirac}}$ has been taken as a reference point, i.e., $I_{\rm{D,p/n}}=I_{\rm{D}}\vert_{V_{\rm{GS}}=V_{\rm{Dirac}}\mp \SI{0.5}{\volt}}$, where the subindexes p and n indicate that the data is taken at the p-type ($V_{\rm{GS}}<V_{\rm{Dirac}}$) or n-type ($V_{\rm{GS}}>V_{\rm{Dirac}}$) part of the transfer curve. The contact resistance $R_{\rm{C}}$ has been extracted \cite{PacFei20} at the corresponding linear unipolar region of the transistor operation.

\vspace{-0.4cm}
\begin{table} [!htb] 
\begin{center}
\caption{Transistor performance indicators at $V_{\rm{DS}}=\SI{0.3}{\volt}$ of the \SI{300}{\nano\meter} GFET using different characterization techniques}
\begin{tabular}{c|c|c||c|c}

 & \multicolumn{2}{c||}{staircase} & \multicolumn{2}{c}{opposing} \\ \hline \hline
param. & \makecell{forward\\sweep} & \makecell{backward\\sweep} & \makecell{forward\\sweep} & \makecell{backward\\sweep} \\ \hline 

$V_{\rm{Dirac}}$ $(\SI{}{\volt})$ & \SI{0.8}{} & \SI{0.9}{} & \SI{0.9}{} & \SI{0.9}{} \\

$I_{\rm{D,p}}/w_{\rm{g}}$ $(\SI{}{\micro\ampere/\micro\meter})$ & \SI{31.45}{} & \SI{31.41}{} & \SI{29.04}{} & \SI{29.25}{} \\


$R_{\rm{C,p}}\cdot w_{\rm{g}}$ $(\SI{}{\kilo\ohm\cdot\micro\meter})$ & \SI{6.61}{} & \SI{7.57}{} & \SI{6.97}{} & \SI{6.71}{} \\

$I_{\rm{D,n}}/w_{\rm{g}}$ $(\SI{}{\micro\ampere/\micro\meter})$ & \SI{28.28}{} & \SI{29}{} & \SI{28.53}{} & \SI{28.36}{} \\

$R_{\rm{C,n}}\cdot w_{\rm{g}}$ $(\SI{}{\kilo\ohm\cdot\micro\meter})$ & \SI{12.35}{} & \SI{13.89}{} & \SI{13.5}{} & \SI{13.1}{} \\

\end{tabular} \label{tab:rc_exp}
\end{center}
\end{table}
\vspace{-0.2cm}

The shielding of the channel potential due to traps leads to an offset in $V_{\rm{GS}}$, i.e., the same current level is obtained at different applied vertical fields in staircase sweeps. Trap-reduced data enables the reproducibility of these carrier current levels despite the measurement history. Larger differences in the contact resistivity from trap-affected data in comparison to the same parameter obtained with opposing pulses are due to the consideration of a bias range rather than a bias point related to the $R_{\rm{C}}$-extraction method \cite{PacFei20}, i.e., hysteresis in this value is scaled up. A wrong estimation of this parameter could mislead the technology development.

\section{Modeling-enabled analog performance assessment}

The need for an analytical current-voltage (\textit{IV}) model that will accurately predict the behavior of both trap-reduced and trap-affected data is crucial for better comprehending these phenomena. Such a model is proposed in this section based on the one derived in \cite{JimMol11} and on the experimental empirical observations presented above. The basic electrostatic equation of a one-gated graphene transistor without the impact of traps leads to an expression for the net charge in graphene given by \cite{JimMol11}, \cite{ThiSch10}

\begin{equation}
Q_{\rm{net}}(x) = C_{\rm{ox}} \left(V_{\rm{GS}}-V_{\rm{GSO}}-V_{\rm{c}}(x)-V_{\rm{ch}}(x)\right),
\label{eq:charge}
\end{equation}

\noindent where $V_{\rm{c}}$ and $V_{\rm{ch}}$ are the chemical and the channel potential at channel position $x$, respectively. The chemical potential accounts for the voltage drop across the quantum capacitance along the channel. $V_{\rm{GS}}-V_{\rm{GSO}}$ is the gate voltage overdrive with $V_{\rm{GSO}}$ used as a model parameter and $C_{\rm{ox}}$ is the oxide capacitance per unit area. By considering a straightforward approach \cite{JimMol11} in which $Q_{\rm net}$ approaches to \SI{0}{} at the charge neutrality point and the channel potential is averaged over the channel such as $V_{\rm{ch}}\sim V_{\rm{DS}}/2$, the Dirac voltage is calculated here as ${\textstyle V_{\rm{Dirac}}\approx V_{\rm{GSO}}+V_{\rm{DS}}/2}$ \cite{JimMol11}. 

\begin{figure}[!htb]
\centering
\includegraphics[height=0.2185\textwidth]{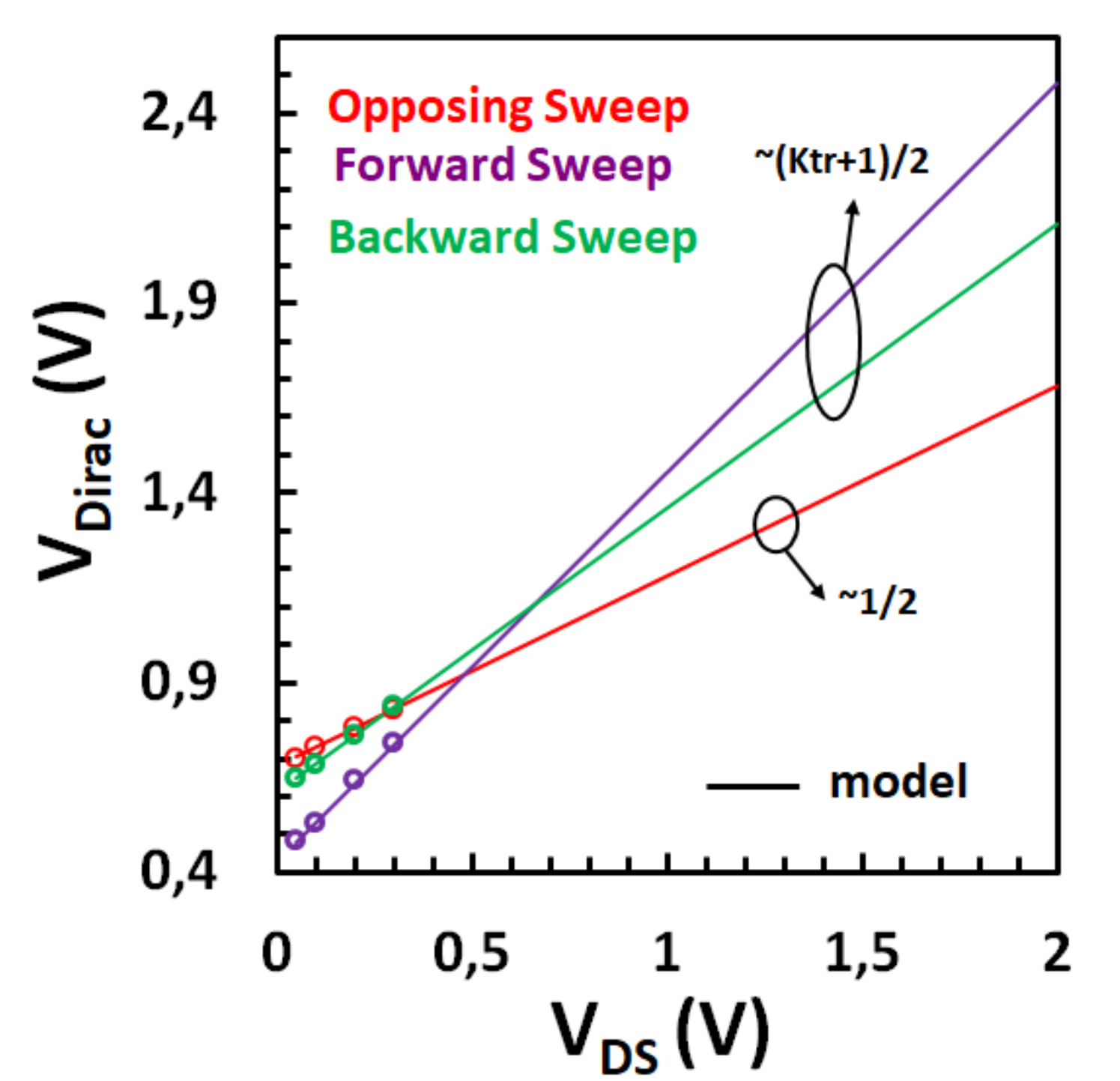}
\includegraphics[height=0.2185\textwidth]{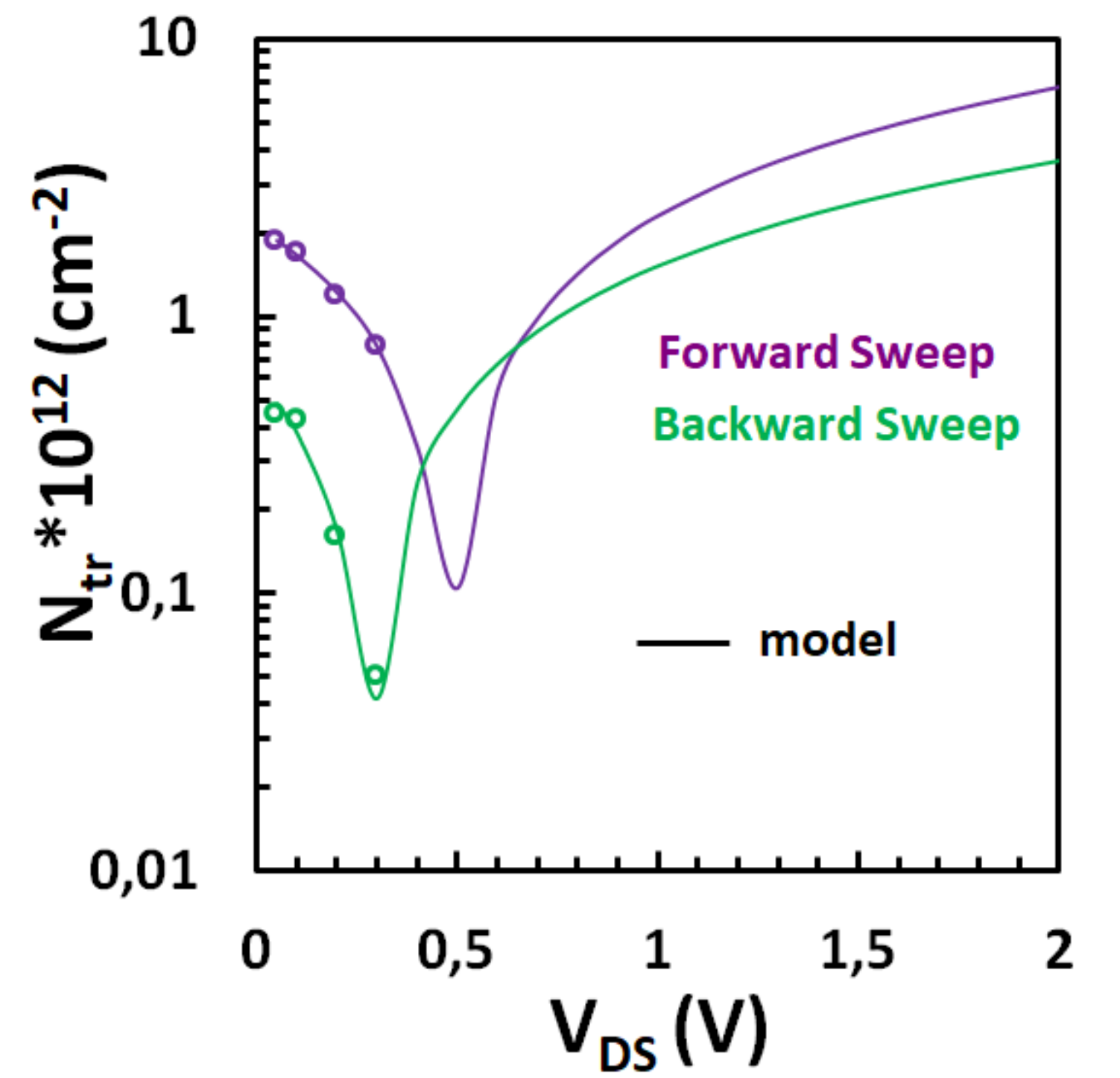} \\
\center{\hspace{0.75cm}(a) \hspace{3.75cm} (b)}
\caption{Measured and simulated Dirac voltage (a) and absolute value of the trap density (b) versus $V_{\rm{DS}}$ for all available sweeps. Markers: data, lines: model. The different slopes of the plots in (a) are pointed out.}
\label{fig:vdirac_Nt}
\end{figure}

At the presence of traps, Eq. (\ref{eq:charge}) is modified as

\begin{equation}
Q_{\rm{net,tr}}(x) = C_{\rm{ox}} \left(V_{\rm{GS}}-V_{\rm{GSO,tr}}-V_{\rm{c}}(x)-V_{\rm{ch}}(x)\right),
\label{eq:charge_02}
\end{equation}

\noindent with ${\textstyle V_{\rm{GSO,tr}}=V_{\rm{GSO}}-DV_{\rm{tr}}+K_{\rm{tr}}V_{\rm{DS}}/2}$. The term $DV_{\rm{tr}}-K_{\rm{tr}}V_{\rm{DS}}/2$ corresponds to the induced potential due to trap density\footnote{Absolute values are considered. See results and discussions of Fig. \ref{fig:vdirac_Nt}(b)} ${\textstyle N_{\rm{tr}}=C_{\rm{ox}}(DV_{\rm{tr}}-K_{\rm{tr}}V_{\rm{DS}}/2)/q}$ with $q$ as the electron charge. $DV_{\rm{tr}}$ and $K_{\rm{tr}}$ are the new defined model parameters where the first corresponds to the shift of $V_{\rm{Dirac}}$ due to trap impact while the latter embraces the $V_{\rm{DS}}$ dependence of this shift. Trap-affected $V_{\rm{Dirac}}$ is now calculated as ${\textstyle V_{\rm{Dirac,tr}}=V_{\rm{GSO}}-DV_{\rm{tr}}+(K_{\rm{tr}}+1)V_{\rm{DS}}/2}$. In order to incorporate the above effects in the $I_{\rm{D}}$ model, $V_{\rm{GSO,tr}}$ is used instead of $V_{\rm{GSO}}$  in Eqs. (1) and (4) of \cite{JimMol11} for the potential calculation. In contrast to other compact GFET model including the impact of traps \cite{FreMen12}, a straightforward implementation of the $V_{\rm{DS}}$-dependence of a trap-affected Dirac voltage related to hot carriers \cite{CarSer14}, \cite{ChiPer10} has been considered here. The latter is an attractive feature to describe the impact of traps in GFET-based circuits at different bias. 

The model accurately captures experimental $I_{\rm{D}}$ and $g_{\rm{m}}$ data as illustrated in Figs. \ref{fig:IdVg_all} and \ref{fig:gmVg_all} for all available sweeps in the p-type region. The asymmetric experimental n-type behavior with respect to the p-type region is not described by the model due to an inherent symmetric condition as discussed elsewhere \cite{MavWei19}. In the following, the discussion is for the transistor operating in the p-type region. The extracted model parameters are shown in Table \ref{tab:iv_param} where $\mu$ is the mobility, $\rho_0$ the residual charge density and $\hbar\Omega$ the phonon energy. The impact of the traps mostly contributes to the shift of the $V_{\rm{Dirac}}$ through $DV_{\rm{tr}}$ and $K_{\rm{tr}}$ parameters which is more intense at the forward sweep while the rest of the model parameters are the same for both trap-reduced and trap-affected data. Having both data sets available, permits a reliable parameter extraction procedure. In more detail, from the opposing sweep all the model parameters apart from $\hbar\Omega$ can be extracted from low $V_{\rm{DS}}$ regime while the precise fitting of the data at high $V_{\rm{DS}}$ by just adjusting $\hbar\Omega$ is a good indicator that impact of the traps is negligible; in such case $DV_{\rm{tr}}$ and $K_{\rm{tr}}$ are deactivated. The latter are extracted afterwards in forward and backward sweeps.

\vspace{-0.4cm}
\begin{table} [!htb] 
\begin{center}
\caption{Model parameters for different measurement sweeps}
\begin{tabular}{c||c|c|c}

parameter &  forward & backward & opposing \\ \hline 

$\mu$ $(\SI{}{\centi\meter^2\volt^{-1}\second^{-1}})$ & \SI{135}{} & \SI{135}{} & \SI{135}{} \\

$V_{\rm{GSO}}$ $(\SI{}{\volt})$ & \SI{0.68}{} & \SI{0.68}{} & \SI{0.68}{} \\

$\rho_0$ $(\SI{}{\centi\meter^{-2}})$ & \SI{1.55e12}{} & \SI{1.55e12}{} & \SI{1.55e12}{} \\

$R_{\rm{C}}/2$ $(\SI{}{\ohm})$ & \SI{116}{} & \SI{116}{} & \SI{116}{} \\

$\hbar\Omega$ $(\SI{}{\milli\electronvolt})$ & \SI{10}{} & \SI{10}{} & \SI{10}{} \\

$DV_{\rm{tr}}$ $(\SI{}{\milli\volt})$ & \SI{250}{} & \SI{70}{} & \SI{0}{} \\

$K_{\rm{tr}}$ & \SI{1.05}{} & \SI{0.5}{} & \SI{0}{} 

\end{tabular} \label{tab:iv_param}
\end{center}
\end{table}
\vspace{-0.2cm}

Fig. \ref{fig:vdirac_Nt}(a) depicts both experimental and simulated $V_{\rm{Dirac}}$ for trap-reduced and trap-affected cases versus $V_{\rm{DS}}$. The model is extended to higher $V_{\rm{DS}}$ values where after a point, $V_{\rm{Dirac}}$ of forward and backward sweeps become larger than the one of opposing case. This can be justified in terms of the different $V_{\rm{DS}}$ dependence of trap-reduced and trap-affected channel potential. When traps are present, this dependence is $(K_{\rm{tr}}+1)V_{\rm{DS}}/2$ instead of $V_{\rm{DS}}/2$ in the case they are not there, as shown before. Experimental trap density is calculated as $N_{\rm{tr}}=\Delta V_{\rm{Dirac}}C_{\rm{ox}}/q$ where $\Delta V_{\rm{Dirac}}$ is the shift of the measured $V_{\rm{Dirac}}$ such as ${\textstyle \Delta V_{\rm{Dirac}}=V_{\rm{Dirac-opp}}-V_{\rm{Dirac-fwd/bwd}}}$, respectively while $N_{\rm{tr}}$ can be also calculated by the model as mentioned previously. Both are illustrated in Fig. \ref{fig:vdirac_Nt}(b) versus $V_{\rm{DS}}$ where the trapping and detrapping processes can be elucidated for voltages lower and higher than the minimum point of the curve, respectively. The minimum point of the $N_{\rm{tr}}(V_{\rm{DS}})$ plot corresponds to a change of polarity of the term $DV_{\rm{tr}}-K_{\rm{tr}}V_{\rm{DS}}/2$. 

The model results in Fig. \ref{fig:vdirac_Nt} confirm the increasing impact of traps with $V_{\rm DS}$ on device performance ellucidated by experimental data, specially on $g_{\rm d}$ (see Fig. \ref{fig:gdVg_all}). Trapping and detrapping processes can be simultaneously enabled at the same $V_{\rm DS}$, e.g., at \SI{0.4}{\volt}, where traps are still being filled during a forward $V_{\rm{GS}}$ sweep while in the backward sweep other traps can either start to release carriers or been enhanced by hot carriers \cite{CarSer14}, \cite{ChiPer10} as suggested by the increase of the modeled $N_{\rm{tr}}$ at this bias point. An accurate calculation and projection over bias of $N_{\rm{tr}}$ such as the one provided by this model (cf. Fig. \ref{fig:vdirac_Nt}(b)), is an important information towards improved insights in the device physics, e.g., on the low-frequency noise characterization of GFETs \cite{MavWei19}.

\begin{figure}[!htb]
\centering
\includegraphics[height=0.2325\textwidth]{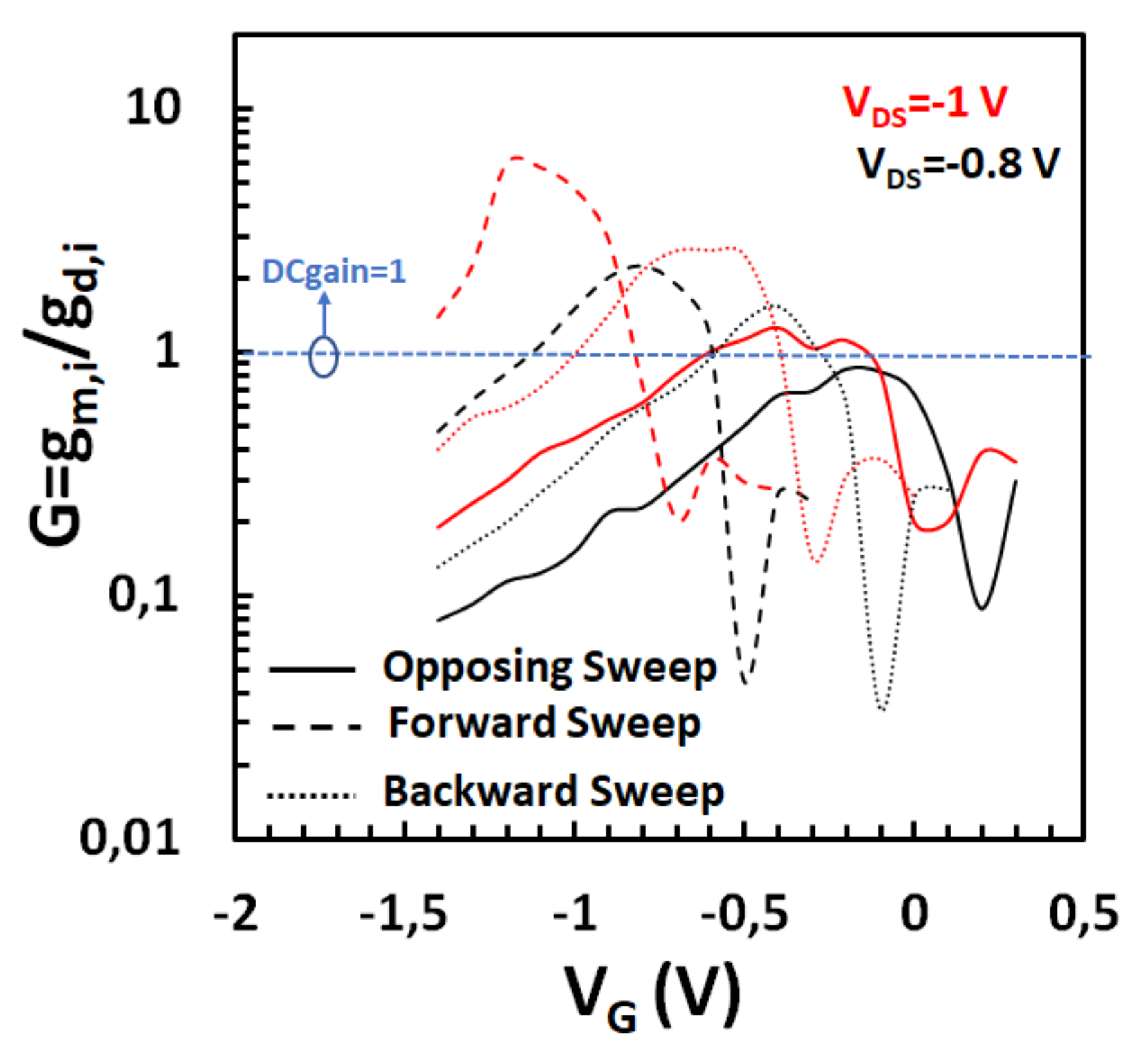}
\caption{Simulated intrinsic gain at different bias for all available sweeps.}
\label{fig:gain}
\end{figure}

Synthetic intrinsic gain $G=g_{\rm{m,i}}/g_{\rm{d,i}}$, calculated with the simulated data\footnote{$V_{\rm GS,i}=V_{\rm GS}-I_{\rm D}R_{\rm C}/2$ and $V_{\rm DS,i}=V_{\rm DS}-I_{\rm D}R_{\rm C}$ with $R_{\rm{C}}$ values of Table \ref{tab:iv_param}.}, is depicted in Fig. \ref{fig:gain} for different $V_{\rm{GS}}$ at the p-type region for two high negative $V_{\rm{DS}}$ values in order to ensure both the p-type operation of the device and that $G$ exceeds unity. It is evident that traps can affect tremendously this important analog FoM as well by affecting the DC bias point and consequently resulting in a \textit{false-positive} result, i.e., an overestimated trap-affected $G$ that can not be exploited in circuit applications since it depends on the measurement history of the device. A more reliable result is the one obtained with the diminished impact of traps which can be reproduced and used in circuits by using an appropiate biasing scheme. 

\begin{figure}[!htb]
\centering
\includegraphics[height=0.2185\textwidth]{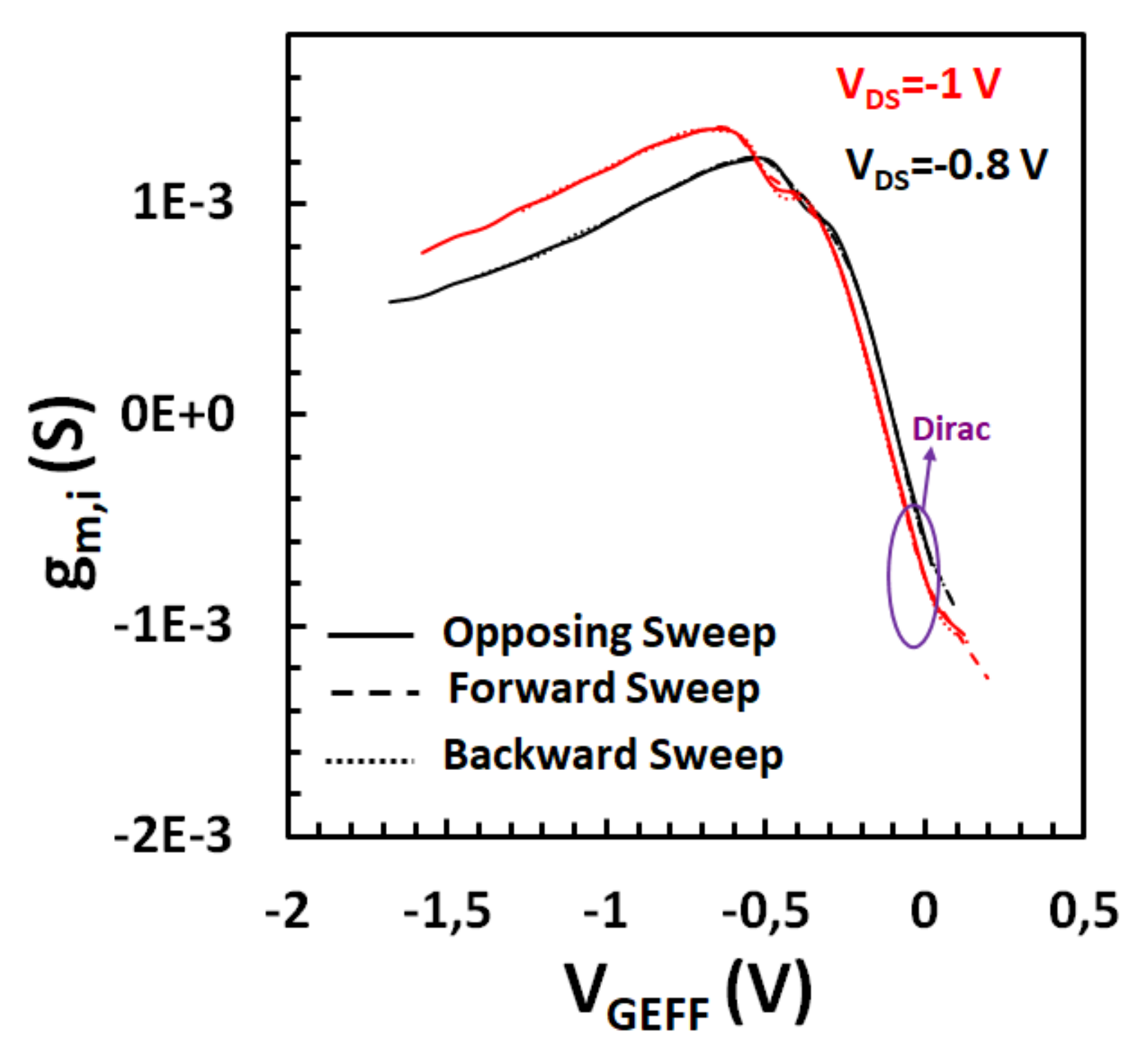}
\includegraphics[height=0.2185\textwidth]{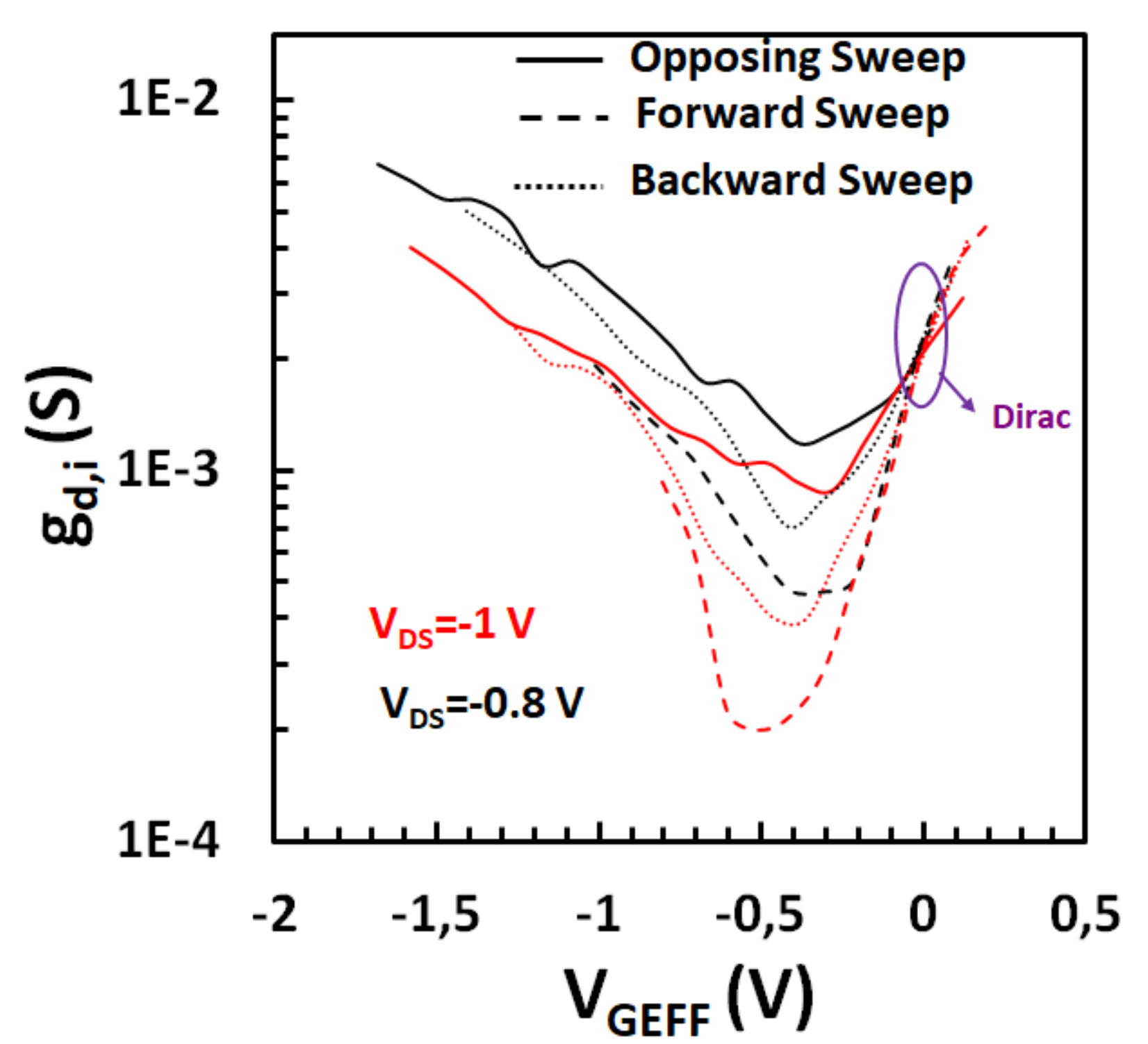} \\
\center{\hspace{0.75cm}(a) \hspace{3.75cm} (b)}
\caption{Simulated intrinsic (a) transconductance and (b) output conductance versus an effective gate voltage $V_{\rm{GEFF}}$.}
\label{fig:gmi_gdi}
\end{figure}

The higher impact of traps on $g_{\rm{d}}$ observed in the experiments (cf. Fig. \ref{fig:gdVg_all}) is also confirmed with the model since the larger the $V_{\rm{DS}}$, the more trapping/detrapping mechanisms can be enabled (higher $N_{\rm{tr}}$). In contrast, the trap-affected $I_{\rm{D}}$ changes slowly with the vertical applied fields, i.e., $g_{\rm{m}}$ is mainly shifted in $V_{\rm{GS}}$ but not in magnitude ($N_{\rm{tr}}$ shifted in $V_{\rm{GS}}$). This is confirmed with the simulated $g_{\rm{m,i}}$ and $g_{\rm{d,i}}$ plots at different bias shown in Fig. \ref{fig:gmi_gdi}.

The effect of traps can be claimed to be diminished regardless the characterization technique if the difference of $V_{\rm{Dirac}}$ with $V_{\rm{GS}}$ is compensated, i.e., by a $V_{\rm{GEFF}}=V_{\rm{GS}}-V_{\rm{Dirac,tr}}$ for the staircase sweeps and $V_{\rm{GEFF}}=V_{\rm{GS}}-V_{\rm{Dirac}}$ for the opposing sweeps. At $V_{\rm{GEFF}}=\SI{0}{\volt}$ the values of the simulated $g_{\rm{m,i}}$ and $g_{\rm{d,i}}$ are the same for all sweeps at the same $V_{\rm{DS}}$ as shown in Fig. \ref{fig:gmi_gdi}. This implies similar $G$ at such bias. However, the condition of identical trap-affected characteristics required to exploit this feature is challenging to reproduce in practical scenarios since they are strongly affected by the measurement history. Hence, trap-reduced data should be always considered for feasible and reproducible device characteristics.

\section{Conclusion}

Consecutive opposite bias pulses applied to a fabricated graphene transistor have been used here in order to diminish the impact of trapping processes on the device performance. This biasing scheme is more practical than the holding time approach generally used to obtain trap-reduced data. In contrast to other GFET studies focused on the device response to varying lateral field in the channel, the opposing pulses stressing the gate of the device here have enabled the observation of trapping effects due mainly to available states in the oxide, i.e., the channel potential is shielded from the applied vertical field due to traps. Trap-affected and trap-reduced static characteristics have been experimentally observed and qualitatively discussed. The oftenly overlooked impact of traps in analog/HF figures of merit has been shown here with calculations of the cutoff frequency and maximum oscillation frequency based on experimental data. The dynamic response can differ from the expected due to the DC bias-point drift induced by trapping processes. Trap-reduced data obtained with practical bias schemes, e.g., by the opposing pulse technique, can ease the design of matching networks for GFETs in high-frequency circuits by providing reproducible characteristics.

An analytical drain current model has been able to reproduce trap-reduced and trap-affected data by considering the impact of traps on the device electrostatics. Two empirical factors have been obtained to describe the $V_{\rm{Dirac}}$ shift and its $V_{\rm{DS}}$ dependence for the trap-affected data. The straightforward implementation of the latter effect in the compact model enables the study of GFET-based circuits including the trap-affected device performance. The trap density obtained from experimental data has been also reproduced by using the model parameters. The synthetic trap density over bias reveals also the effect of trapping and detrapping processes. An important analog figure of merit such as the intrinsic gain has been obtained here from the simulated data. Similarly to the HF FoMs ($f_{\rm t}$ and $f_{\rm max}$), an important difference is observed for this analog metric between trap-affected and trap-reduced results. Traps affect not only the biasing point but the magnitude of the intrinsic gain. 

The impact of traps should be considered in both static and dynamic scenarios in order not to spoil the analog/HF performance of GFETs. A practical and straightforward biasing scheme as the one used here as well as the modeling approach presented in this work are convenient for both obtaining the trap-reduced device performance and exploiting the device dynamic characteristics in circuits.

\end{document}